\documentclass[twoside]{article}
\usepackage{graphicx}

\begin{document}

\centerline{\Large \bf 
Three-Quark Potential in SU(3) Lattice QCD 
}

\vspace{0.75cm}

\centerline{
T.T. Takahashi$^1$, H. Matsufuru$^1$, Y. Nemoto$^2$ and H. Suganuma$^{3, 1}$}

\vspace{0.5cm}

\centerline{\it $^1$ RCNP, Osaka University, Mihogaoka 10-1, Osaka 567-0047, Japan}
\centerline{\it $^2$ YITP, Kyoto University, Kitashirakawa, Sakyo, Kyoto 606-8502, Japan}
\centerline{\it $^3$ Faculty of Science, 
        Tokyo Institute of Technology, Tokyo 152-8551, Japan}

\vspace{0.3cm}
      
\begin{abstract}

The static three-quark (3Q) potential is studied in SU(3) 
lattice QCD with $12^3 \times 24$ and $\beta=5.7$ at the 
quenched level. From the 3Q Wilson loop, 
3Q ground-state potential $V_{\rm 3Q}$ is extracted 
using the smearing technique for ground-state enhancement. 
With accuracy better than a few \%, $V_{\rm 3Q}$ is well 
described by a sum of a constant, the two-body Coulomb term 
and the three-body linear confinement term $\sigma_{\rm 3Q} L_{\rm min}$, 
with $L_{\rm min}$ the minimal value of total length 
of color flux tubes linking the three quarks. 
Comparing with the Q-$\bar {\rm Q}$ potential, 
we find a universal feature of the string tension,  
$\sigma_{\rm 3Q} \simeq \sigma_{\rm Q \bar Q}$,  
and the OGE result for Coulomb 
coefficients, $A_{\rm 3Q} \simeq \frac12 A_{\rm Q \bar Q}$. 
\end{abstract}

\vspace{0.3cm}


In usual, the three-body force is regarded as a residual 
interaction in most fields of physics.  
In QCD, however, the three-body force among three quarks  
is expected to be a ``primary'' force reflecting the SU(3)$_c$ gauge symmetry. 
Indeed, the three quark (3Q) potential \cite{SMNT00,BPV95,RS91} 
is directly responsible to the structure and properties 
of baryons \cite{CI86}, similar to the relevant role of 
the Q-$\bar {\rm Q}$ potential upon meson properties \cite{LSG91}. 
In contrast with a number of studies on the Q-$\bar {\rm Q}$ 
potential using lattice QCD \cite{BS92,K00}, 
there were only a few lattice QCD studies for the 3Q potential  
done mainly more than 13 years ago \cite{SW8486,KEFLM87,TES88,B00}. 
(In Ref.\cite{B00}, the author only showed a preliminary result 
on the equilateral-triangle case without enough analyses.)
In Refs.\cite{SW8486,TES88,B00},  
the 3Q potential seemed to be expressed by a sum of two-body potentials, 
which supports the $\Delta$-type flux tube picture \cite{C96}.
On the other hand, Ref.\cite{KEFLM87} seemed to support 
the Y-type flux-tube picture \cite{BPV95,CI86} rather than 
the $\Delta$-type one. 
These controversial results may be due to the difficulty of the 
accurate measurement of the 3Q ground-state potential in lattice QCD. 
For instance, in Refs.\cite{SW8486,TES88}, the authors did not 
use the smearing for ground-state enhancement, and therefore their results  
may include serious contamination from the excited-state component.  

The 3Q static potential can be measured with the 3Q Wilson loop, 
where the 3Q gauge-invariant state 
is generated at $t=0$ and is annihilated at $t=T$, 
as shown in Fig.1. Here, the three quarks are spatially fixed 
in ${\bf R}^3$ for $0 < t < T$. 
The 3Q Wilson loop $W_{\rm 3Q}$ is defined in a gauge-invariant manner as 
\begin{equation}
W_{\rm 3Q} \equiv \frac1{3!}\varepsilon_{abc}\varepsilon_{a'b'c'}
U_1^{aa'} U_2^{bb'} U_3^{cc'} 
\end{equation}
with $U_k \equiv {\rm P}\exp\{ig\int_{\Gamma_k}dx^\mu A_{\mu}(x)\}$ 
($k=1,2,3$).  
Here, $P$ denotes the path-ordered product along the path denoted by 
${\Gamma_k}$ in Fig.1.
Similar to the derivation of the Q-${\bar {\rm Q}}$ potential 
from the Wilson loop, the 3Q potential $V_{\rm 3Q}$ is obtained as 
$V_{\rm 3Q}=-\lim_{T \rightarrow \infty} \frac1T 
\ln \langle W_{\rm 3Q}\rangle.$

Physically, the true ground state of the 3Q system, which is of interest here, 
is expected to be expressed by the flux tubes instead of the strings, 
and the 3Q state which is expressed by the three strings generally includes 
many excited-state components such as flux-tube vibrational modes. 
Of course, if the large $T$ limit can be taken, the 
ground-state potential would be obtained. However, 
$\langle W_{\rm 3Q}\rangle$ decreases exponentially with $T$, 
and then the practical measurement of $\langle W_{\rm 3Q}\rangle$ 
becomes quite severe for large $T$ 
in lattice QCD simulations. 
Therefore, for the accurate measurement of the 3Q ground-state 
potential $V_{\rm 3Q}$, the smearing technique for
ground-state enhancement \cite{SMNT00,APE87,BSS95} 
is practically indispensable. 
However, this smearing technique was not applied to 
the past lattice QCD studies for $V_{\rm 3Q}$ in Refs. \cite{SW8486,TES88}, 
since the smearing technique were mainly developed after their works.

In this paper, we study the 3Q ground-state potential $V_{\rm 3Q}$ 
using the ground-state enhancement by the gauge-covariant 
smearing technique for the link-variable 
in SU(3)$_c$ lattice QCD with the standard action with 
$\beta$=5.7 and $12^3 \times$ 24 
at the quenched level \cite{SMNT00}. 
We consider 16 patterns of the 3Q configuration where the three quarks are 
put on $(i,0,0)$, $(0,j,0)$ and $(0,0,k)$ in ${\bf R}^3$ with 
$0 \le i,j,k \le 3$ in the lattice unit. 
Here, the junction point $O$ is set at the origin $(0,0,0)$ in ${\bf R}^3$, 
although the final result of the ground-state potential 
$V_{\rm 3Q}$ should not depend on the artificial selection of $O$. 
For actual lattice QCD calculations of the 3Q Wilson loop,
we use the translational, the rotational and the reflection symmetries
on lattices on the choice of the origin $O$ and 
the direction of $\hat{x},\hat{y},\hat{z}$


The standard smearing for link-variables is expressed as the 
iteration of the replacement of the spatial link-variable $U_i (s)$ 
($i=1,2,3$) by the obscured link-variable 
$\bar U_i (s) \in {\rm SU(3)}_c$ \cite{APE87,BSS95} which maximizes  
\begin{equation}
{\rm Re} \,\,{\rm tr} \left\{ \bar U_i^{\dagger}(s) \left[
\alpha U_i(s)+\sum_{j \ne i} \{
U_j(s)U_i(s+\hat j)U_j^\dagger (s+\hat i) + 
U_j^\dagger (s-\hat j)U_i(s-\hat j)U_j(s+\hat i-\hat j)
\} \right] \right\}
\end{equation}
with the smearing parameter $\alpha$ being a real number. 
The $n$-th smeared link-variables $U_\mu^{(n)}(s)$ $(n=1,2,..,N_{\rm smear})$ 
are iteratively defined starting from $U_\mu^{(0)}(s) \equiv U_\mu(s)$ as
\begin{equation}
U_i^{(n)}(s) \equiv \bar U_i^{(n-1)}(s) 
\quad 
(i=1,2,3), 
\qquad 
U_4^{(n)}(s) \equiv U_4(s).
\end{equation}

As an important feature, this smearing procedure keeps 
the gauge covariance of the ``fat'' link-variable $U_\mu^{(n)}(s)$ properly. 
In fact, the gauge-transformation property of 
$U_\mu^{(n)}(s)$ is just the same as that of the original link-variable 
$U_\mu(s)$, and therefore 
the gauge invariance of $F(U_\mu^{(n)}(s))$ is ensured 
whenever $F(U_\mu (s))$ is a gauge-invariant function. 

Since the fat link-variable $U_\mu^{(n)}(s)$ includes a spatial 
extension, the ``line'' expressed with $U_\mu^{(n)}(s)$ physically 
corresponds to a ``flux tube'' with a spatial extension.
Therefore, if a suitable smearing is done, the ``line'' of the 
fat link-variable is expected to be close to the ground-state flux tube. 
(This smearing technique is actually successful for the extraction of 
the Q-$\bar {\rm Q}$ potential in lattice QCD \cite{BSS95}.) 

Now, we investigate the magnitude of the ground-state component in  
the 3Q-state operator at $t=0, T$ in the 3Q Wilson loop $W_{\rm 3Q}$.    
From the similar argument in the Q-$\bar{\rm Q}$ system \cite{BSS95}, 
the overlap of the 3Q-state operator with the ground state 
is estimated with 
\begin{equation}
C_0 \equiv \langle W_{\rm 3Q} (T) \rangle ^{T+1} / 
\langle W_{\rm 3Q} (T+1) \rangle ^T, 
\end{equation}
which we call the ground-state overlap. 
If the 3Q state at $t=0, T$ in Fig.1 is the perfect ground state, 
$\langle W_{\rm 3Q} (T) \rangle = e^{-V_{\rm 3Q} T}$ 
and then $C_0$=1 hold.  
(Here, $W_{\rm 3Q} (T)$ in Eq.(1)  is normalized as  
$\langle W_{\rm 3Q} (T=0) \rangle=1$.) 

The ground-state potential $V_{\rm 3Q}$ can be measured accurately 
if $C_0$ is large enough. Then, we check the ground-state overlap $C_0$ of 
the 3Q Wilson loop $\langle W_{\rm 3Q} (U_\mu^{(n)}(s), T) \rangle$
composed of the fat link-variable $U_\mu^{(n)}(s)$ 
in lattice QCD simulations, and search reasonable values of 
the smearing parameter $\alpha$ and the iteration number 
$N_{\rm smear}$ for this purpose. 
In lattice QCD simulations, the ground-state overlap $C_0$ 
is largely enhanced as $0.8 < C_0 < 1$ for $T \le 3$ 
by the smearing with $\alpha=2.3$ and $N_{\rm smear}=12$ 
for all of the 3Q configurations in consideration, 
as shown in Fig.2. 


Here, we make a theoretical consideration on the potential form 
in the Q-$\bar{\rm Q}$ and 3Q systems with respect to QCD. 
In the short-distance limit, 
perturbative QCD is applicable and  
the Coulomb-type potential appears 
as the one-gluon-exchange (OGE) result. 
In the long-distance limit at the quenched level,
the flux-tube picture would be applicable from the argument of 
the strong-coupling limit of QCD \cite{BPV95,CI86,KS75},  
and hence a linear-type confinement potential 
proportional to the total flux-tube length 
is expected to appear. 
Of course, it is nontrivial that these simple arguments on the 
ultraviolet and infrared limits of QCD hold for the 
intermediate region as $0.2 \,{\rm fm} < r < 1 \,{\rm fm}$. 
Nevertheless, lattice QCD results for 
the Q-$\bar {\rm Q}$ ground-state potential is well described by 
\begin{equation}
V_{\rm Q \bar{Q}}(r)=-\frac{A_{\rm Q \bar{Q}}}{r}
+\sigma_{\rm Q \bar{Q}} r+C_{\rm Q \bar{Q}}  
\end{equation}
at the quenched level \cite{BSS95}.  
Actually, we measure $V_{\rm Q \bar{Q}}$ from the on-axis Wilson loop 
with the smearing with $\alpha=2.3$ and $N_{\rm smear}=20$, and find 
a good fitting of Eq.(5) with the parameters 
($A_{\rm Q \bar{Q}}$, $\sigma_{\rm Q \bar{Q}}$, $C_{\rm Q \bar{Q}}$) 
listed in Table~1. 
(In general, the adequate values of $\alpha$ and $N_{\rm smear}$ 
depend on the operator.)
In fact, $V_{\rm Q \bar{Q}}$ is described by a sum of the short-distance 
OGE result and the long-distance flux-tube result. 

Also for the 3Q ground-state potential $V_{\rm 3Q}$, we try to 
apply this simple picture of the short-distance OGE result 
plus the long-distance flux-tube result. 
Then, the 3Q potential $V_{\rm 3Q}$ is expected to take a form of 
\begin{equation}
V_{\rm 3Q}=-A_{\rm 3Q}\sum_{i<j}\frac1{|{\bf r}_i-{\bf r}_j|}
+\sigma_{\rm 3Q} L_{\rm min}+C_{\rm 3Q}, 
\end{equation}
where $L_{\rm min}$ denotes the minimal value of 
total length of color flux tubes linking the three quarks.
Denoting three sides of the 3Q triangle by $a$,$b$ and $c$, 
$L_{\rm min}$ is expressed as 
\begin{equation}
L_{\rm min}=\left[\frac12 (a^2+b^2+c^2)
 +\frac{\sqrt{3}}2 
  \sqrt{(a+b+c)(-a+b+c)(a-b+c)(a+b-c)}\right]^{\frac12}, 
\end{equation}
when all angles of the 3Q triangle do not exceed $2\pi/3$.
In this case, there appears the physical junction 
which connects the three flux tubes originating from the three quarks, 
and the shape of the 3Q system is expressed as a Y-type flux 
tube \cite{CI86,KS75}, where the angle between two flux tubes is found 
to be $2\pi/3$ \cite{BPV95,CI86}. 
When an angle of the 3Q triangle exceeds $2\pi/3$, 
one finds $L_{\rm min}=a+b+c-{\rm max}(a,b,c)$.


Now, we show lattice QCD results for the 3Q ground-state potential. 
We generate 210 gauge configurations using SU(3)$_c$ lattice QCD 
Monte-Carlo simulation with the standard action with 
$\beta=5.7$ and $12^{3} \times 24$ at the quenched level.
The lattice spacing $a \simeq 0.19 \,{\rm fm}$ is determined 
so as to reproduce the string tension as $\sigma$=0.89 GeV/fm
in the Q-$\bar{\rm Q}$ potential $V_{\rm Q \bar{Q}}$.
Here, the gauge configurations are taken every 500 sweeps 
after a thermalization of 5000 sweeps using the pseudo-heat-bath algorithm.
In this study, lattice QCD calculations have been performed on NEC-SX4
at Osaka University.  

We measure the 3Q ground-state potential $V_{\rm 3Q}$ 
using the smearing technique, and 
compare the lattice data with the theoretical form of Eq.(6).
Owing to the smearing, the ground-state component is largely enhanced,  
and therefore the 3Q Wilson loop $\langle W_{\rm 3Q} \rangle$ 
composed with the smeared link-variable exhibits 
a single-exponential behavior as 
$\langle W_{\rm 3Q} \rangle \simeq e^{-V_{\rm 3Q}T}$ 
even for a small value of $T$. 
Then, for each 3Q configuration, we extract $V_{\rm 3Q}^{\rm latt}$
from the least squares fit with the single-exponential form
\begin{equation}
\langle W_{\rm 3Q}\rangle =\bar{C}e^{-V_{\rm 3Q}T}
\end{equation}
in the range of $T_{\rm min}\leq T\leq T_{\rm max}$ listed in Table~2.
Here, we choose the fit range of $T$ such that the stability of the
``effective mass''
$V(T)\equiv \ln\{\langle W_{\rm 3Q}(T) \rangle /
\langle W_{\rm 3Q}(T+1)\rangle\}$
is observed in the range of $T_{\rm min}\leq T\leq T_{\rm max}-1$.
%

For each 3Q configuration, 
we summarize the lattice data $V_{\rm 3Q}^{\rm latt}$ 
as well as the prefactor $\bar{C}$ in Eq.(8), 
the fit range of $T$ and $\chi^2/N_{\rm DF}$ in Table~2. 
The statistical error of $V_{\rm 3Q}^{\rm latt}$ is estimated 
with the jackknife method. 
We find again a large ground-state overlap as $\bar{C} > 0.8$ 
for all 3Q configurations.

Now, we consider the potential form. 
In Fig.3, we plot the 3Q ground-state potential $V_{\rm 3Q}$    
as the function of $L_{\rm min}$.
Apart from a constant, $V_{\rm 3Q}$  is almost proportional 
to $L_{\rm min}$ in the infrared region. 
We show in Table~1 the best fitting parameters in Eq.(6) for $V_{\rm 3Q}$.
On the goodness of this fitting, $\chi^2$ seems relatively large 
as $\chi^2/N_{\rm DF}=3.76$, 
which may reflect a systematic error on the finite lattice spacing.   

We add in Table~2 
the comparison of the lattice data $V_{\rm 3Q}^{\rm latt}$ with 
the fitting function $V_{\rm 3Q}^{\rm fit}$ 
in Eq.(6) with the parameters listed in Table~1. 
The three-quark ground-state potential $V_{\rm 3Q}$ is 
well described by Eq.(6) with accuracy better than a few \%.

Next, we compare the coefficients 
$(\sigma_{\rm 3Q}, A_{\rm 3Q}, C_{\rm 3Q})$ 
in the 3Q potential $V_{\rm 3Q}$ in Eq.(6) with 
$(\sigma_{\rm Q\bar{Q}}, A_{\rm Q\bar{Q}}, C_{\rm Q\bar{Q}})$ 
in the Q-$\bar {\rm Q}$ potential $V_{\rm Q\bar{Q}}$
in Eq.(5) as listed in Table~1. 
As a remarkable fact, we find a universal feature of the string tension, 
$\sigma_{\rm 3Q} \simeq \sigma_{\rm Q\bar{Q}}$,  
as well as the OGE result for the Coulomb coefficient, 
$A_{\rm 3Q} \simeq \frac12 A_{\rm Q\bar{Q}}$. 

As a check, we consider the diquark limit, where two quark 
locations coincide in the 3Q system. 
In the diquark limit, the static 3Q system becomes equivalent to 
the Q-$\bar{\rm Q}$ system, 
which leads to a physical requirement on the relation between 
$V_{\rm 3Q}$ and $V_{\rm Q \bar{Q}}$. 
Our results, $\sigma_{\rm 3Q} \simeq \sigma_{\rm Q\bar{Q}}$ and 
$A_{\rm 3Q} \simeq \frac12 A_{\rm Q\bar{Q}}$, 
are consistent with the physical requirement in the diquark limit. 
Here, the constant term is to be considered carefully in the diquark limit, 
because there appears a singularity or a divergence 
from the Coulomb term in $V_{\rm 3Q}$ in the continuum diquark limit, 
$\lim_{{\bf r}_j \rightarrow {\bf r}_i} 
\frac{-A_{\rm 3Q}}{|{\bf r}_i-{\bf r}_j|}=-\infty$. 
In the lattice regularization, this ultraviolet divergence 
is regularized to be a finite constant with the lattice spacing $a$ as 
$\frac{-A_{\rm 3Q}}{|{\bf r}_i-{\bf r}_j|} \rightarrow 
\frac{-A_{\rm 3Q}}{\omega a}$, where $\omega$ is a dimensionless constant 
satisfying $0 < \omega <1$ and $\omega \sim 1$. 
Then, we find 
$C_{\rm 3Q}+\frac{-A_{\rm 3Q}}{\omega a}=C_{\rm Q\bar{Q}}$, i.e.,  
$C_{\rm 3Q}-C_{\rm Q\bar{Q}}=\frac{A_{\rm 3Q}}{\omega a} \quad ( > 0) $
in the diquark limit.  
This is the requirement for the constant term 
in the diquark limit on the lattice. 
Our lattice QCD results for $C_{\rm 3Q}$, $C_{\rm Q \bar{Q}}$ 
and $A_{\rm 3Q}$ are thus consistent with this requirement,  
and one finds $\omega \simeq 0.46$. 

Finally, we also try to fit $V_{\rm 3Q}^{\rm latt}$ with 
the $\Delta$-type flux-tube ansatz of  
$V_{\rm 3Q}=-A_\Delta \sum_{i<j}\frac1{|{\bf r}_i-{\bf r}_j|}
+\sigma_\Delta \sum_{i<j} |{\bf r}_i-{\bf r}_j|+C_\Delta$,
which was suggested in Refs.\cite{SW8486,TES88,C96}.
However, this fitting seems rather worse, because $\chi^2$ is unacceptably 
large as $\chi^2/N_{\rm DF}=10.1$ even for the best fit with 
$A_\Delta=0.1410(64)$, $\sigma_\Delta=0.0858(16)$ and 
$C_\Delta=0.9344(210)$ in the lattice unit.
As an approximation, $V_{\rm 3Q}$ seems described by a simple sum of 
the effective two-body Q-Q potential with a reduced string tension as 
$\sigma_{\rm QQ} \simeq 0.53  \sigma$. 
This reduction factor can be naturally understood as a geometrical factor 
rather than the color factor, since 
the ratio between $L_{\rm min}$ and the perimeter length $L_P$ satisfies 
$\frac12 \le \frac{L_{\rm min}}{L_P} \le \frac1{\sqrt{3}}$, which 
leads to $L_{\rm min} \sigma = L_P \sigma_{\rm QQ}$ 
with $\sigma_{\rm QQ}=(0.5 \sim 0.58) \sigma$. 

We have studied the static 3Q ground-state potential in SU(3) lattice QCD 
at the quenched level, using the smearing technique for
ground-state enhancement. 
We have found that $V_{\rm 3Q}$ is well described 
by a sum of a constant, the two-body Coulomb term and the three-body 
linear confinement term $\sigma_{\rm 3Q} L_{\rm min}$, 
where  $L_{\rm min}$ denotes the minimal length of the color flux tube 
linking the three quarks. 
We have also found a universal feature of the string tension as   
$\sigma_{\rm 3Q} \simeq \sigma_{\rm Q \bar Q}$,  
and the OGE result for Coulomb 
coefficients as $A_{\rm 3Q} \simeq \frac12 A_{\rm Q \bar Q}$. 
In lattice QCD studies, however, there appear systematic errors 
relating to the finite lattice-spacing effect and so on.
To obtain the conclusive result on the 3Q potential, we are investigating 
with finer lattices with large $\beta$'s.

\begin{figure}[p]
\begin{center}
\includegraphics[height=6cm]{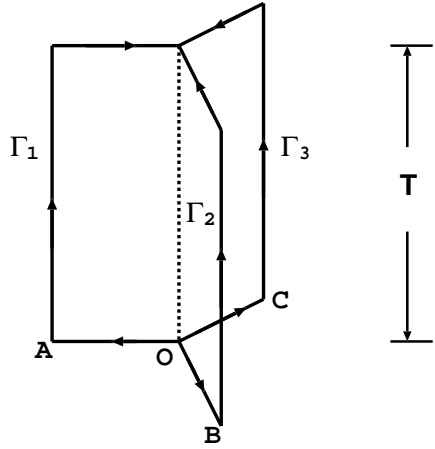}
\caption{The 3Q Wilson loop $W_{\rm 3Q}$.
The 3Q state is generated at $t=0$ and is annihilated at $t=T$. 
The three quarks are spatially fixed in ${\bf R}^3$ for $0 < t < T$.
}
\end{center}
\end{figure}

\begin{figure}[p]
\begin{center}
\includegraphics[height=7cm]{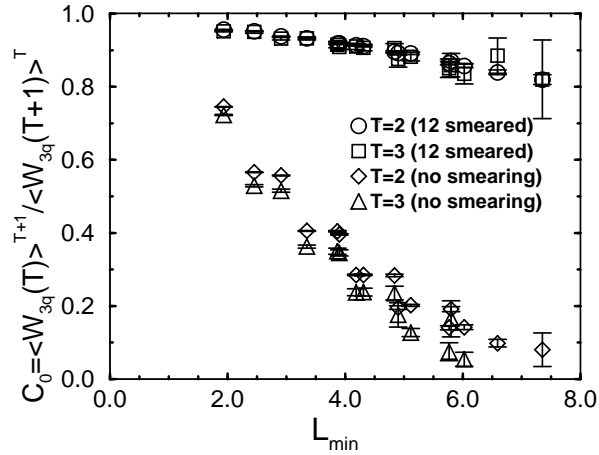}
\vspace{-0.5cm}
\caption{The ground-state overlap of the 3Q system, 
$C_0 \equiv \langle W_{\rm 3Q} (T) \rangle ^{T+1} / 
\langle W_{\rm 3Q} (T+1) \rangle ^T$,  
with the smeared link-variable (upper data) 
and with unsmeared link-variable (lower data). 
To distinguish the 3Q system, we have taken   
the horizontal axis as $L_{\rm min}$, which the minimal length 
of the flux tubes linking the three quarks. 
For each 3Q configuration, 
$C_0$ is largely enhanced as $0.8 < C_0 < 1$ by the smearing. 
}
\end{center}
\end{figure}

\begin{figure}[p]
\begin{center}
\includegraphics[height=8cm]{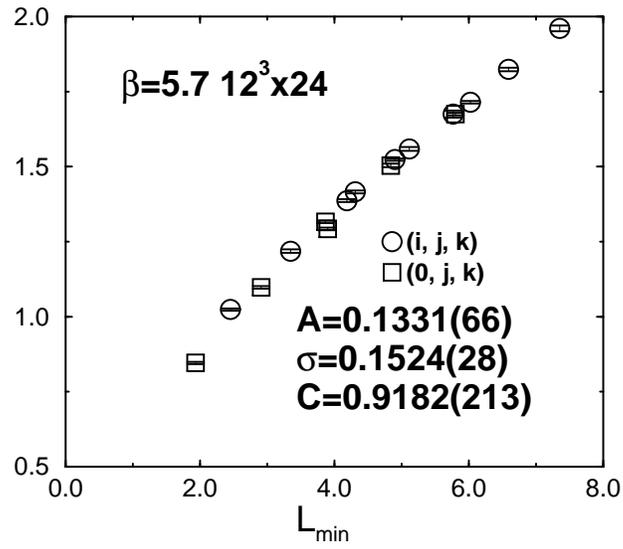}
\caption{
The 3Q ground-state potential $V_{\rm 3Q}$   
as the function of $L_{\rm min}$ which is the minimum value of 
the total length of the flux tubes. 
}
\end{center}
\end{figure}

\begin{table}[htb]
\begin{center}
\caption{
The coefficients in Eq.(6) for the 3Q potential and 
those in Eq.(5) for the Q-$\bar {\rm Q}$ potential in the lattice unit.
}
\vspace{0.2cm}
\newcommand{\m}{\hphantom{$-$}}
\newcommand{\cc}[1]{\multicolumn{1}{c}{#1}}
\renewcommand{\tabcolsep}{2pc} 
\renewcommand{\arraystretch}{1.2} 
\begin{tabular}{@{}llll} \hline \hline
                   & \cc{$\sigma$} & \cc{$A$}      & \cc{$C$}     \\ \hline
${\rm 3Q}$       & $0.1524(28)$  & $0.1331( 66)$  & $0.9182(213)$  \\ 
${\rm Q\bar{Q}}$ & $0.1629(47)$  & $0.2793(116)$ & $0.6203(161)$ \\ 
\hline \hline
\end{tabular}\\[2pt]
\end{center}
\end{table}


\begin{table}[htb]
\caption{
Lattice QCD results for the 3Q potential $V_{\rm 3Q}^{\rm latt}$
for 16 patterns of the 3Q system, where the three quarks are put on $(i,0,0)$, 
$(0,j,0)$ and $(0,0,k)$ in ${\bf R}^3$ in the lattice unit.
For each 3Q configuration, $V_{\rm 3Q}^{\rm latt}$ is measured 
from the single-exponential fit 
$\langle W_{\rm 3Q}\rangle=\bar{C}e^{-V_{\rm 3Q}T}$ 
in the range of $T$ listed at the fourth column. 
The statistical errors listed are estimated with the jackknife method,  
and $\chi^2/N_{\rm DF}$ is listed at the fifth column. 
The fitting function $V_{\rm 3Q}^{\rm fit}$ in Eq.(6) with 
the best fitting parameters in Table~1 is also added. 
}
\vspace{0.2cm}
\newcommand{\m}{\hphantom{$-$}}
\newcommand{\cc}[1]{\multicolumn{1}{c}{#1}}
\begin{center}
\begin{tabular}{ c c c c c c c } \hline\hline
$(i, j, k)$ & $V_{\rm 3Q}^{\rm latt}$ & \lower.4ex\hbox{$\bar{C}$} &  
fit range of $T$ & $\chi^2/N_{\rm DF}$
& $V_{\rm 3Q}^{\rm fit}$ & $V^{\rm latt}_{\rm 3Q}-V^{\rm fit}_{\rm 3Q}$ \\
\hline
$(0, 1, 1)$ &0.8457( 38)& 0.9338(173)& 5 --10 & $0.062$ & 0.8524& $-$0.0067 \\ 
$(0, 1, 2)$ &1.0973( 43)& 0.9295(161)& 4 -- 8 & $0.163$ & 1.1025& $-$0.0052 \\ 
$(0, 1, 3)$ &1.2929( 41)& 0.8987(110)& 3 -- 7 & $0.255$ & 1.2929&  \m0.0000 \\
$(0, 2, 2)$ &1.3158( 44)& 0.9151(120)& 3 -- 6 & $0.053$ & 1.3270& $-$0.0112 \\ 
$(0, 2, 3)$ &1.5040( 63)& 0.9041(170)& 3 -- 6 & $0.123$ & 1.5076& $-$0.0036 \\ 
$(0, 3, 3)$ &1.6756( 43)& 0.8718( 73)& 2 -- 5 & $0.572$ & 1.6815& $-$0.0059 \\ 
$(1, 1, 1)$ &1.0238( 40)& 0.9345(149)& 4 -- 8 & $0.369$ & 1.0092&  \m0.0146 \\ 
$(1, 1, 2)$ &1.2185( 62)& 0.9067(228)& 4 -- 8 & $0.352$ & 1.2151&  \m0.0034 \\ 
$(1, 1, 3)$ &1.4161( 49)& 0.9297(135)& 3 -- 7 & $0.842$ & 1.3964&  \m0.0197 \\ 
$(1, 2, 2)$ &1.3866( 48)& 0.9012(127)& 3 -- 7 & $0.215$ & 1.3895& $-$0.0029 \\ 
$(1, 2, 3)$ &1.5594( 63)& 0.8880(165)& 3 -- 6 & $0.068$ & 1.5588&  \m0.0006 \\ 
$(1, 3, 3)$ &1.7145( 43)& 0.8553( 76)& 2 -- 6 & $0.412$ & 1.7202& $-$0.0057 \\ 
$(2, 2, 2)$ &1.5234( 37)& 0.8925( 65)& 2 -- 5 & $0.689$ & 1.5238& $-$0.0004 \\ 
$(2, 2, 3)$ &1.6750(118)& 0.8627(298)& 3 -- 6 & $0.115$ & 1.6763& $-$0.0013 \\ 
$(2, 3, 3)$ &1.8239( 56)& 0.8443( 90)& 2 -- 5 & $0.132$ & 1.8175&  \m0.0064 \\
$(3, 3, 3)$ &1.9607( 93)& 0.8197(154)& 2 -- 5 & $0.000$ & 1.9442&  \m0.0165 \\ 
\hline\hline
\end{tabular}\\[2pt]
\end{center}
\end{table}

\begin{thebibliography}{9}
\bibitem{SMNT00} 
H. Suganuma, H. Matsufuru, Y. Nemoto and T.T. Takahashi, 
Nucl. Phys. {\bf A680} (2000) 159.
\bibitem{BPV95} N. Brambilla, G.M. Prosperi and A. Vairo, 
Phys. Lett. {\bf B362} (1995) 113. 
\bibitem{RS91} M. Fable de la Ripelle and Yu. A. Simonov, Ann. Phys. 
{\bf 212} (1991) 235.
\bibitem{CI86} S. Capstick and N. Isgur, Phys. Rev. {\bf D34} (1986) 2809. 
\bibitem{LSG91} 
W. Lucha, F.F.Sch\"oberl and D.Gromes, Phys. Rep. {\bf 200} (1991) 127.
\bibitem{BS92} G.S. Bali and K. Schilling, 
Phys. Rev. {\bf D46} (1992) 2636. 
\bibitem{K00} K. Schilling,  
Nucl. Phys. {\bf B} (Proc.Suppl.) {\bf 83-84} (2000) 140 and 
references therein.
\bibitem{SW8486} R. Sommer and J. Wosiek, 
Phys. Lett. {\bf 149B} (1984) 497; Nucl. Phys. {\bf B267} (1986) 531.
\bibitem{KEFLM87} J. Kamesberger et al., 
Proc of ``Few-Body Problems in Particle, 
Nuclear, Atomic and Molecular Physics'' (1987) 529. 
\bibitem{TES88} H.B. Thacker, E. Eichten and J.C. Sexton, 
Nucl. Phys. {\bf B} (Proc. Suppl.) {\bf 4} (1988) 234.
\bibitem{B00} G.S. Bali, preprint hep-ph/0001312 (2000).
\bibitem{C96} J.M. Cornwall, Phys. Rev. {\bf D54} (1996) 6527. 
\bibitem{APE87} APE Collaboration, M. Albanese et al., 
Phys. Lett. {\bf B192} (1987) 163. 
\bibitem{BSS95} G.S. Bali, C. Schlichter and K. Schilling, 
Phys. Rev. {\bf D51} (1995) 5165. 
\bibitem{KS75} J. Kogut and L. Susskind, Phys. Rev. {\bf D11} (1975) 395. 

\end{thebibliography}
\end{document}